\documentclass[multphys,vecphys]{svmult}

\usepackage{makeidx}         
\usepackage{graphicx}        
\usepackage{multicol}        
\usepackage[bottom]{footmisc}
\usepackage{natbib}

\makeindex             

\begin{document}

\title*{Tracing the Mass--Assembly History of Galaxies with Deep Surveys}
\author{Georg Feulner\inst{1,2},
  Armin Gabasch\inst{1,2},
  Yuliana~Goranova\inst{1,2},
  Ulrich Hopp\inst{1,2},
  \and Ralf Bender\inst{1,2}}

\authorrunning{Georg Feulner et al.}

\institute{Universit\"ats--Sternwarte M\"unchen, Scheinerstra\ss e 1,
D--81679 M\"unchen, Germany, \texttt{feulner@usm.lmu.de} \and
Max--Planck Institut f\"ur extraterrestrische Physik,
Giessenbachstra\ss e 1, D--85748 Garching, Germany}
%
%
\maketitle

\begin{abstract}
We use the optical and near-infrared galaxy samples from the Munich
Near-Infrared Cluster Survey (MUNICS), the FORS Deep Field (FDF) and
GOODS-S to probe the stellar mass assembly history of field galaxies
out to $z \sim 5$. Combining information on the galaxies' stellar mass
with their star-formation rate and the age of the stellar
population, we can draw important conclusions on the assembly of the
most massive galaxies in the universe: These objects contain the
oldest stellar populations at all redshifts probed. Furthermore, we
show that with increasing redshift the contribution of star-formation
to the mass assembly for massive galaxies increases dramatically,
reaching the era of their formation at $z \sim 2$ and beyond. These
findings can be interpreted as evidence for an early epoch of star
formation in the most massive galaxies in the universe.
\end{abstract}

\section{Introduction}

In recent years, there has been considerable interest in the relation
of the stellar mass in galaxies and their star-formation rate (SFR),
since this allows to quantify the contribution of the recent star
formation to the build up of stellar mass for different galaxy masses.
\citet{feulner:Cowie1996} were the first to investigate this
connection for a $K$-selected sample of $\sim 400$ galaxies at $z<1.5$
and noted an emerging population of massive, heavily star forming
galaxies at higher redshifts, a phenomenon they termed
``down-sizing''.  Later on, the specific star-formation rate (SSFR),
defined as the SFR per unit stellar mass, was used to study this
relation.

\section{Connecting Star Formation and Stellar Mass}

\begin{figure}
\centering
\includegraphics[height=0.75\textwidth,angle=270]{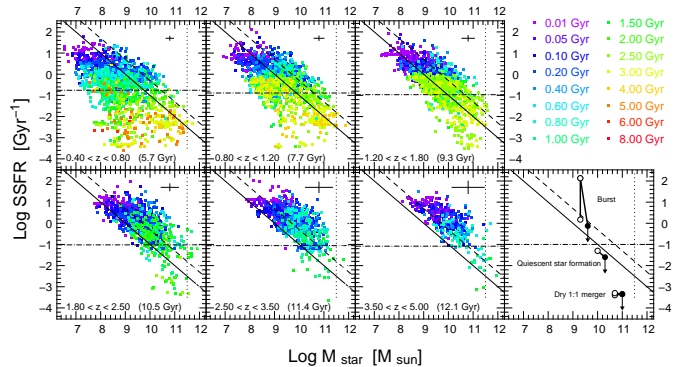}
\caption{\textit{Six panels to the left:} The SSFR as a function of
  stellar mass and redshift. The solid and dashed lines correspond to
  constant SFRs. Objects are coloured according to the age of their
  stellar population. The dot-dashed line is the SSFR required to
  double a galaxy's mass between each redshift epoch and today
  (assuming constant SFR); the corresponding look-back time is
  indicated as well. The error cross in each panel gives an idea of
  the typical errors, while the dotted line roughly represents the
  high-mass cut-off of the local stellar mass function
  \citep{feulner:munics6, feulner:fdfmf}. \textit{Lower right-hand
  panel:} Examples for evolutionary paths yielding a doubling of a
  galaxy's mass. Open symbols denote the starting point, filled
  symbols the final state; the arrows indicated the influence of gas
  consumption or loss.}
\label{feulner:fig1}
\end{figure}

We have analysed the SSFR as a function of stellar mass and redshift
$z$ out to $z = 1.2$ \citep{feulner:munics7} using a large sample of
more than 6000 $I$-band selected galaxies from MUNICS
\citep{feulner:munics1, feulner:munics5}. The SSFR decreases with mass
at all redshifts, although we might not detect highly obscured
galaxies. The low values of the SSFR of the most massive galaxies
suggests that most of these massive systems formed the bulk of their
stars at earlier epochs. Furthermore, stellar population synthesis
models show that the most massive systems contain the oldest stellar
populations at all redshifts. This is in agreement with the detection
of old, massive galaxies at redshifts $1 < z < 2$
\citep{feulner:tesis1, feulner:Cimatti2004, feulner:tesis3} and beyond
\citep{feulner:ChenMarzke2004}. In Fig.~\ref{feulner:fig1}, where we
have used the FDF \citep{feulner:fdf1, feulner:fdflf1} and GOODS-S
samples, we show that this trend continues to even higher redshifts
\citep{feulner:fdfssfr}.

\section{The Build-up of the Most Massive Galaxies}

\begin{figure}
\centering
\includegraphics[width=0.45\textwidth]{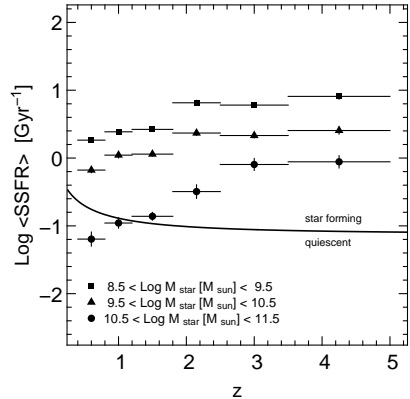}
\caption{Average SSFRs for galaxies with stellar masses of $\log
M_\star/M_\odot \in [8.5,9.5]$ (squares), $[9.5,10.5]$ (triangles) and
$[10.5,11.5]$ (circles) and SFRs larger than $1 \: M_\odot \:
\mathrm{yr}^{-1}$ as a function of $z$. The error bar represents the
error of the mean. The solid line indicates the doubling line of
Fig.~\ref{feulner:fig1} which can be used to discriminate quiescent
and heavily star forming galaxies.}
\label{feulner:fig2}
\end{figure}

It is extremely interesting to investigate the average SSFR of
galaxies with different masses as a function of redshift shown in
Fig.~\ref{feulner:fig2} \citep{feulner:fdfssfr}. While at redshifts $z
< 2$ the most massive galaxies are in a quiescent state, at redshifts
$z > 2$ the SSFR for massive galaxies increases by a factor of $\sim
10$ reaching the epoch of their formation at $z \sim 2$ and
beyond. While there is evidence for dry merging (i.e.\ without
interaction-induced star formation) in the field galaxy population
\citep{feulner:Faber2005, feulner:Bell2005b}, this strong increase in
the SSFR of the most massive galaxies suggests that at least part of
this population was formed in an early period of efficient star
formation in massive haloes.

\printindex
\end{document}